\documentclass[conf]{new-aiaa}
\usepackage[utf8]{inputenc}

\usepackage{graphicx}
\usepackage{amsmath}
\usepackage[version=4]{mhchem}
\usepackage{siunitx}
\usepackage{longtable,tabularx}
\usepackage{todonotes}

\graphicspath{{./figures/}}

\title{Enhancing Aeroacoustic Wind Tunnel Studies through Massive Channel Upscaling with MEMS Microphones}

\author{Daniel Ernst\footnote{Scientist, DLR AS-EXV, Bunsenstrasse 10, 37038 Göttingen, Germany}, Armin Goudarzi*, Reinhard Geisler*, Florian Philipp\footnote{Software Engineer, DLR AS-EXV, Bunsenstrasse 10, 37038 Göttingen, Germany}, Thomas Ahlefeldt*, and Carsten Spehr\footnote{Group Leader, DLR AS-EXV, Bunsenstrasse 10, 37038 Göttingen, Germany}}
\affil{DLR - German Aerospace Center, Göttingen, 37083 Germany}

\begin{document}

\maketitle

\begin{abstract}
This paper presents a large 6~m x 3~m aperture 7200 MEMS microphone array. The array is designed so that sub-arrays with optimized point spread functions can be used for beamforming and thus, enable the research of source directivity in wind tunnel facilities. The total array consists of modular 800 microphone panels, each consisting of four unique PCB board designs. This modular architecture allows for the time-synchronized measurement of an arbitrary number of panels and thus, aperture size and total number of sensors. The panels can be installed without a gap so that the array's microphone pattern avoids high sidelobes in the point spread function. The array's capabilities are evaluated on a 1:9.5 airframe half model in an open wind tunnel at DNW-NWB. The total source emission is quantified and the directivity is evaluated with beamforming. Additional far-field microphones are employed to validate the results. 
\end{abstract}

\section{Introduction}

Microphone array measurements have a long history in aeroacoustic testing and are vital in quantifying acoustic source locations and their sound power of complex objects such as aircraft~\cite{Herkes1998,Oerlemans2004,Soderman2004}) to reduce sound emission levels~\cite{Europe2020,Peters2018}.

Current microphone arrays are limited at low frequencies through the array's aperture and at high frequencies through aliasing due to the microphone spacing. In general, larger apertures will increase the resolution, and more sensors will reduce aliasing and increase the dynamic range. 

In the pursuit of advancing aeroacoustic research, the integration of cutting-edge technologies has become imperative to unravel the complexities of aeroacoustic phenomena. This paper explores a novel approach by employing massive channel upscaling using Micro-Electro-Mechanical Systems (MEMS) microphones within open jet wind tunnel experiments. The goal of this approach is to capture several aeroacoustic properties that are currently neglected such as source directivity to deepen the understating of aerodynamic flows and airframe noise in high-lift configurations.

In open jet wind tunnels the aperture size is typically of the nozzle size since shear layer decorrelation~\cite{ErnstSpehrBerkefeld2015} prevents accurate measurements at higher frequencies and large distances between sensors. So, larger arrays with many more microphones together with standard processing provide only minimal advantages over state-of-the-art ones in industrial wind tunnels.

The main advantage of a large array is the flexibility to perform beamforming with sub-arrays from different incident angles to take the model rotation into account for different Angles of Attack (AoA) or evaluate source directivities. In addition, the high microphone density is capable of fully resolving the spatial sound field enabling complete source identification and quantification approaches. When stating the acoustic source identification problem as an inverse problem~\cite{Raumer2021}, the number of knowns are in the order of $\mathcal{O}(M^2)$ (unique values of the Hermitian Cross Spectral Matrix (CSM) without diagonal), where $M$ are the number of microphones, and the number of unknowns is the number of steering points $N$ in the beamforming map (typically around $10^6$ to $10^7$). In the case of typical microphone arrays ($M\approx$100~\cite{Ahlefeldt2013,Ahlefeldt2017,Ahlefeldt2023}), the number of unknowns exceeds that of knowns, resulting in under-determined problems. With an increase in sensors of a factor of 10 to 100, the inverse problem is getting over-determined.

\begin{figure}[ht!]
\centering
\includegraphics[width=1.0\textwidth]{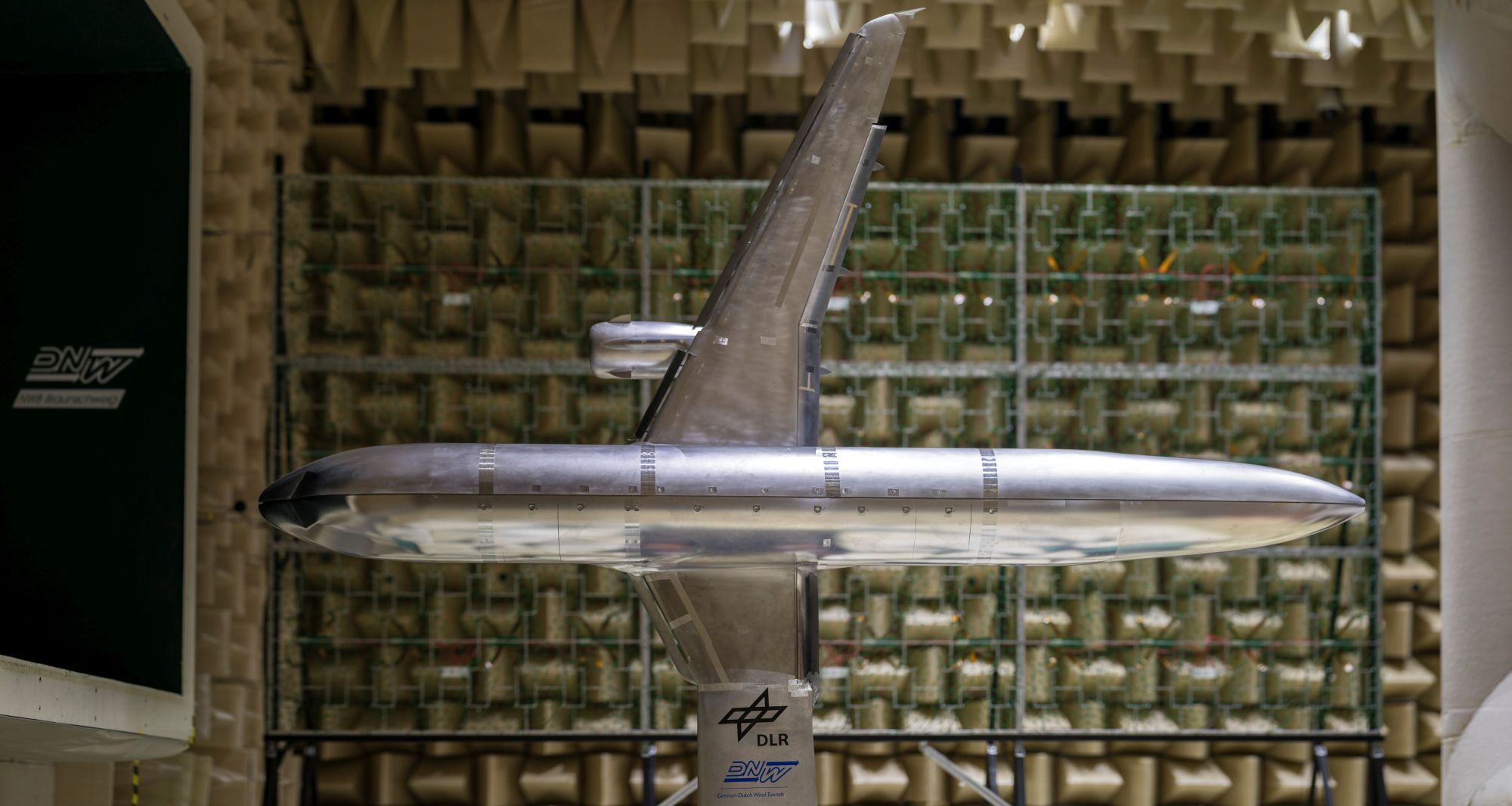}
\caption{Photo of the model and the array in the open test section.}
\label{fig:photo}
\end{figure}

This paper outlines the implementation of the MEMS microphone array, featuring a sensor spacing that allows the use of arbitrary sub-arrays with an optimized Point Spread Function (PSF). Further, the data acquisition and processing are discussed. Last, an open wind tunnel experiment with a 1:9.5 half model, see Figure~\ref{fig:photo}, is presented and evaluated to highlight the benefits of the large MEMS array over conventional arrays.

\section{MEMS Microphone Array}
A novel MEMS microphone data acquisition platform was developed using synchronized Field Programmable Gate Arrays (FPGA) collecting the raw data stream from the sensors and sending it to a host server. Overall, the array has a rectangular aperture of $\SI{6}{\metre}\times\SI{3}{\metre}$ with a total number of 7200 microphones. The microphone array consist of 9 panels ($\SI{1}{\metre}\times\SI{2}{\metre}$). Each panel contains 16 printed circuit boards (PCB), each of which contains 50 microphones. The components of the array are described below.

\subsection{MEMS Sensor}
The array employs Infineon IM69D130 MEMS sensors. They feature a sensitivity accuracy of $\pm\SI{1}{\decibel}$ and phase precision of $\pm\SI{1}{\degree}$ at $\SI{1}{\kilo\hertz}$ based on their data sheet and thus, offer sufficient precision for aeroacoustic research. Their compact dimension ($\SI{4}{\milli\metre}\times\SI{3}{\milli\metre}\times\SI{1.2}{\milli\metre}$) facilitates an easy integration into experimental setups, requiring only a low $\SI{1}{\milli\ampere}$ current consumption. An analog-digital converter is already integrated into the MEMS sensor and the digital data is sent as a Pulse Density Modulated (PDM) 1~bit stream at 3.072~MHz. Over the frequency range of 60~Hz to 6~kHz, the sensor maintains a $\pm\SI{2}{\decibel}$ accuracy in frequency response. The noise floor of 25~dB(A) is comparable to a standard 1/2" condenser microphone. The whole measurement chain including the MEMS Sensor, the mounting, and the data acquisition is calibrated in an anechoic chamber using reference free field microphones of type GRAS~46BF. The phase match is assured by a pressure field calibration using a GRAS~46DE reference microphone and a USound MEMS source.
The calibration of the array is not part of this paper and will be published separately. 

\subsection{PCB Design}
The overall goal was to implement a large microphone array, capturing the acoustics from multiple directions. A common limitation in manufacturing printed circuit boards is the size of the actual machine that assembles the electronic parts on the PCB. Thus, the sensors have to be arranged on multiple PCBs that together form a large array. Common microphone layouts are nonrepetitive patterns such as spirals, however, the main cost driver for low-volume PCB manufacturing and assembly is initial setup costs per layout. 

A compromise between these goals was chosen by using four different PCB layouts of size $\SI{250}{\milli\metre}\times\SI{500}{\milli\metre}$, so that the total array consists of panels with a fixed pattern of $\SI{500}{\milli\metre}\times\SI{1000}{\milli\metre}$. This final pattern has no gaps between PCBs that would introduce PSF side lobes.

The PCB is designed to be as acoustically and aerodynamically transparent as possible. Each MEMS Microphone is placed in the center of a circular area with a 5 mm radius. Small beams of only 3 mm width between the MEMS containing all wires act as mechanical support at the same time.

\begin{figure}[ht!]
\centering
\includegraphics[width=1.0\textwidth]{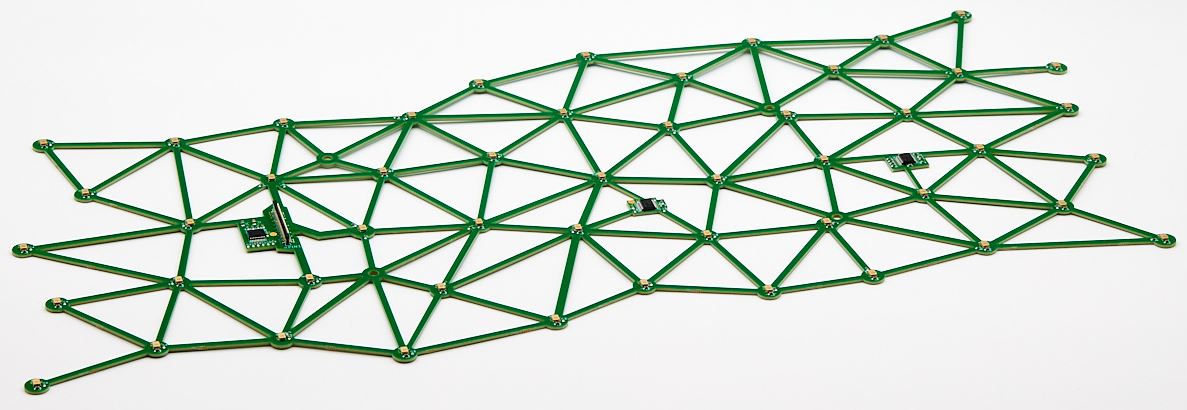}
\caption{Sub-array with 50 MEMS microphones of size $\SI{250}{\milli\metre}\times\SI{500}{\milli\metre}$.}
\label{fig:S10K_OS_UR}
\end{figure}

Figure\ref{fig:S10K_OS_UR} shows one of the four unique PCB designs. Power is supplied over DC cables with 24~V from a single source and is stepped down to 3.3~V at the FPGAs and where it is needed. The overall power consumption does not exceed 120~W.  

\subsection{Data aquisition}
The raw PDM bit streams from 200 MEMS microphones are each collected synchronously by a single FPGA, and are then sent together with meta information such as timestamps, status, etc., as User Datagram Protocol (UDP) network packets to a server at a continuous data rate of approximately $\SI{620}{\mega\bit\per\second}$. Synchronization is assured over a single clock source that is star-shaped distributed to all FPGAs using Low Voltage Differential Signaling (LVDS) over same-length CAT~6 network cables. There are 36 FPGAs involved in this measurement.

The maximum part-to-part skew for the whole chain from the clock source to the FPGAs is 1.2~ns.

The clock distribution from FPGA to the MEMS sensors will add a delay of 1.1~ns. The synchronization between different FPGAs adds no systematic jitter and involves a scheme to detect and report synchronization errors (e.g. due to electromagnetic interference) and ensure a subsequent re-synchronization. In summary, the maximum phase difference of the clock distribution between two arbitrary channels is below $\SI{0.03}{\degree}$ at 20~kHz (360 $\cdot$ 3 ns $\cdot$ 20 kHz $\approx \SI{0.02}{\degree}$ ).\\

All data from the FPGA is written directly to disk on a server, so that measurement periods are only limited by disk size. The raw Pulse Density Modulation (PDM) stream of 1~bit at 3.072~MHz is converted to a 32~bit Pulse Code Modulation (PCM) at 48~kHz by a four-stage decimation filter on the server for further data processing.\\

\section{Open wind tunnel experiment at DNW-NWB}
\begin{figure}[ht!]
\centering
\includegraphics[width=1.0\textwidth]{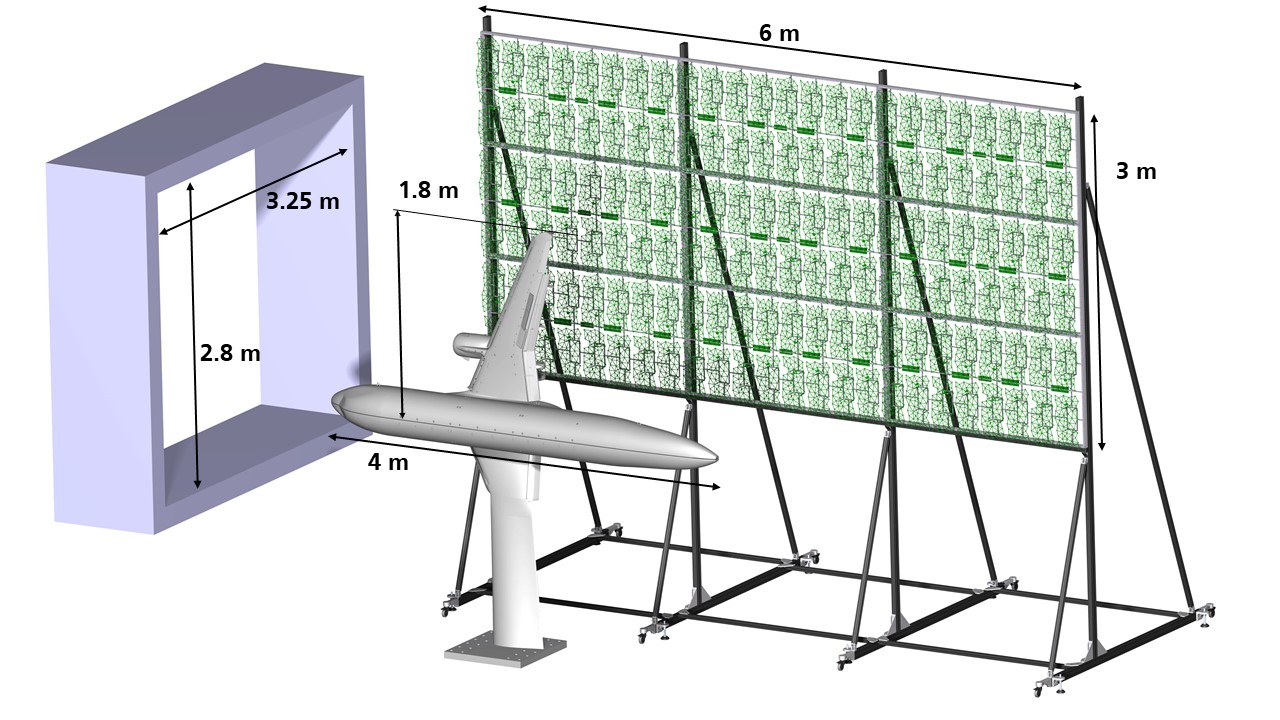}
\caption{Illustration of wind tunnel setup with nozzle, half-model and microphone array. The array is placed $d=\SI{3.39}{\metre}$ away from the model.}
\label{fig:SIAM_Model}
\end{figure}

Reference airframe noise measurements are performed at the low-speed wind tunnel DNW-NWB, Braunschweig, Germany. This facility is a closed-circuit low-speed wind tunnel. In this test the open test section with a nozzle size of $\SI{3.25}{\metre} \times \SI{2.8}{\metre}$ is used. Figure~\ref{fig:SIAM_Model} shows the wind tunnel setup with the nozzle on the left, the model in the center, and the out-of-flow array on the right. 

Inside the test section is a half model of a single-aisle aircraft of scale 1:9.5 in high lift configuration, with movable slat and flap sections and a removable right main gear and nose landing gear. There are replaceable low-noise modifications for the high lift systems and the landing gears. The model surface roughness is below $Ra\le \SI{0.8}{\micro\metre}$. Inside the model are 250 static pressure taps distributed over the slat, wing and flap. The model's aerodynamic and aeroacoustic performance is investigated under AoA $\SI{-6}{\degree}\le\alpha\le\SI{25}{\degree}$, each for a Mach number range $0.1\le\mathrm{M}\le0.20$.

\section{Methods}\label{sec:methods}
This section describes the methods used for the results throughout the paper. Subsection~\ref{sec:method:beamforming} presents the beamforming methods used. Subsection~\ref{sec:method:directivity} presents the methods used to derive source directivities. Subsection~\ref{sec:method:farfield} presents the methods to compare beamforming results and far-field spectra.

\subsection{Beamforming}\label{sec:method:beamforming}
The cross-spectral matrix (CSM) is calculated using Welch's method with a block size of 1024 time samples with \SI{50}{\percent} overlap and a Hanning window. The source emission is estimated by conventional beamforming~\cite{Sijtsma_NLR_2012} with steering vector formulation III~\cite{Sarradj2012}, a monopole assumption, and CLEAN-SC~\cite{Sijtsma2007} with diagonal removal. The speed of sound, convection, and humidity (atmospheric damping) of the fluid is integrated into the Green's function~\cite{Ahlefeldt2023}. Additionally, sound refraction through the shear layer is corrected using the Amiet open wind tunnel correction~\cite{AMIET1978}.

Beamforming is performed on an equidistant 2D focus grid, that follows the wing's delta angle, and the AoA, see Figure~\ref{fig:bfm_grid}. While the resolution of the focus grid is low, it is sufficient to evaluate the overall sound emission of the model. 

The beamforming estimated source level depends on the cross-spectral density, which is affected by effects that reduce the coherence between sensor pairs. The Amiet open wind tunnel correction only assumes refraction along a plane through the shear layer and does not correct for scattering-induced signal decorrelation. This decorrelation depends on the thickness of the shear layer, the angle of incident, the frequency, and the distance between the sensor pairs~\cite{Ernst2020}\cite{ErnstSpehrBerkefeld2015}.

\subsection{Directivity}\label{sec:method:directivity}
Aeroacoustic noise is typically associated with dipole noise, which has a directivity. Directivity in this context means, that more source energy is projected into some direction than others, i.e. the dipole radiates energy to its main lobes, but not perpendicular to them. 

Figure~\ref{fig:SIAM_Model} shows the wind tunnel setup with nozzle on the left, the model in the center, and the out-of-flow array on the right. This allows for observation angles of
$\SI{55}{\degree}\le\theta\le\SI{137}{\degree}$, and $\SI{-27}{\degree}\le\varphi\le\SI{27}{\degree}$ for an exemplary point in the middle of the wing at $\mathbf{x}_0=[2.4,0,0]^T$. Observation angles are given relative to the model so that $\theta$ is the pitch angle and  $\varphi$ is the roll angle. 

The directivity will be calculated based on the geometric mean of a sub-array in relation to the reference point at the wing $\mathbf{x}_0$. The geometric mean of a sub-array is the average location of all sensors. Further, beamforming applies a spatial filter on the observed sound field, so that it is not only observed at the given angle but in the vicinity of the geometric mean, described by the standard deviation of the sensor locations.

Note, that the true directivity depends on the actual location of a source in 3D space. Thus, beamforming must be evaluated in 3D~\cite{Ahlefeldt2023}, individual sources must be identified~\cite{Goudarzi2021}, and the directivity must then be calculated based on the object rotation (which includes the model AoA, but also on the relative object rotation, such as a flap angle). For this first study, these effects are neglected and only the directivity for the total model is assessed.

\subsection{Beamforming to far-field projection}\label{sec:method:farfield}
Beamforming estimates the source power at a given focus point, using a monopole assumption in this case. From the sound power, the resulting sound power immission level at any point in space can be deduced using a monopole Green's function. The monopole assumes a decay of the sound power level by $1/d^2$, where $d$ is the distance to the source.

To compare the results of beamforming, and the far-field microphone, they must be normalized to a sound immission level at a normalized distance. To do so, the beamforming result is spatially integrated and calculated as the sound pressure that would be observed at $d_0=\SI{1}{\metre}$ from the source. The far-field microphones are then corrected with $\Delta \text{PSD}=20\log_{10}(d/d_0)$, where $d_0=\SI{1}{\metre}$ is the reference distance of the beamforming result.

\section{Results}\label{sec:results}

This section presents results for AoA $\alpha=\SI{8.7}{\degree}, \text{M}=0.2$, and it structured as follows. Subsection~\ref{sec:results:sensorvalidation} presents results of the MEMS sensors (from here referred to as ``DLR sensors'') and compares them to the DNW array (140 microphones of type LinearX M51) and far-field microphones (G.R.A.S. 40AC). Far-field microphone results will be displayed in green, DNW array results in blue, and DLR array results in red. Subsection~\ref{sec:results:directivity} presents results from sub-arrays at various pitch angles, to obtain source directivities. Subsection~\ref{sec:results:farfield} presents results of a frequency-dependent aperture array and compares its beamforming performance to the far-field microphones.

All tests were performed twice, once for the DNW array, and once for the simultaneous acquisition with the DLR-array and far-field microphones.

\subsection{Sensor validation}\label{sec:results:sensorvalidation}
\begin{figure}[ht!]
\centering
\includegraphics[width=1.0\textwidth]{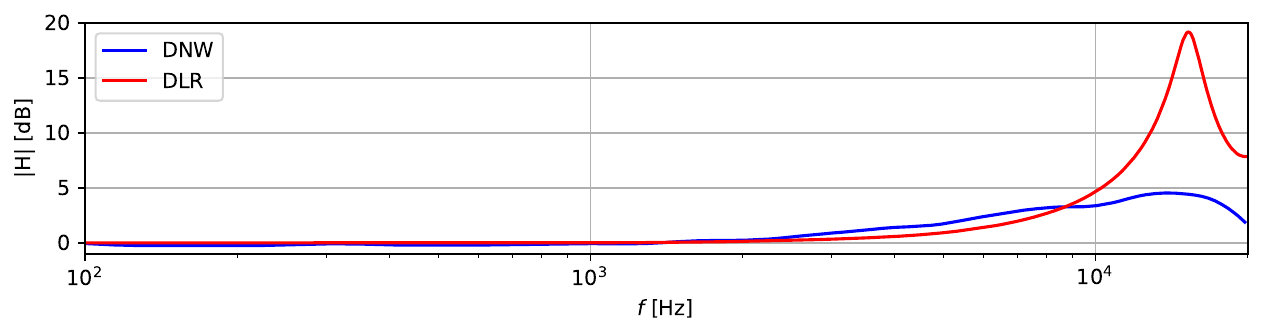}
\caption{Frequency response of the DNW microphones and DLR MEMS.}
\label{fig:freq_response_vgl}
\end{figure}

This subsection presents an in-depth comparison of the DNW, DLR, and far-field microphones. Figure~\ref{fig:freq_response_vgl} shows the sensor frequency response of the DLR and DNW microphones. Note, that the response for the DNW microphones is averaged over all microphones, as they show aging, which shifts the resonance frequencies resulting in $\pm 5 dB$ variation above 10 kHz. The DLR sensors show a defined prominent resonance at $f=\SI{15}{\kilo\hertz}$. This result was obtained with both a free-field calibration and a pressure-field calibration. 

\begin{figure}[ht!]
\centering
\includegraphics[width=1.0\textwidth]{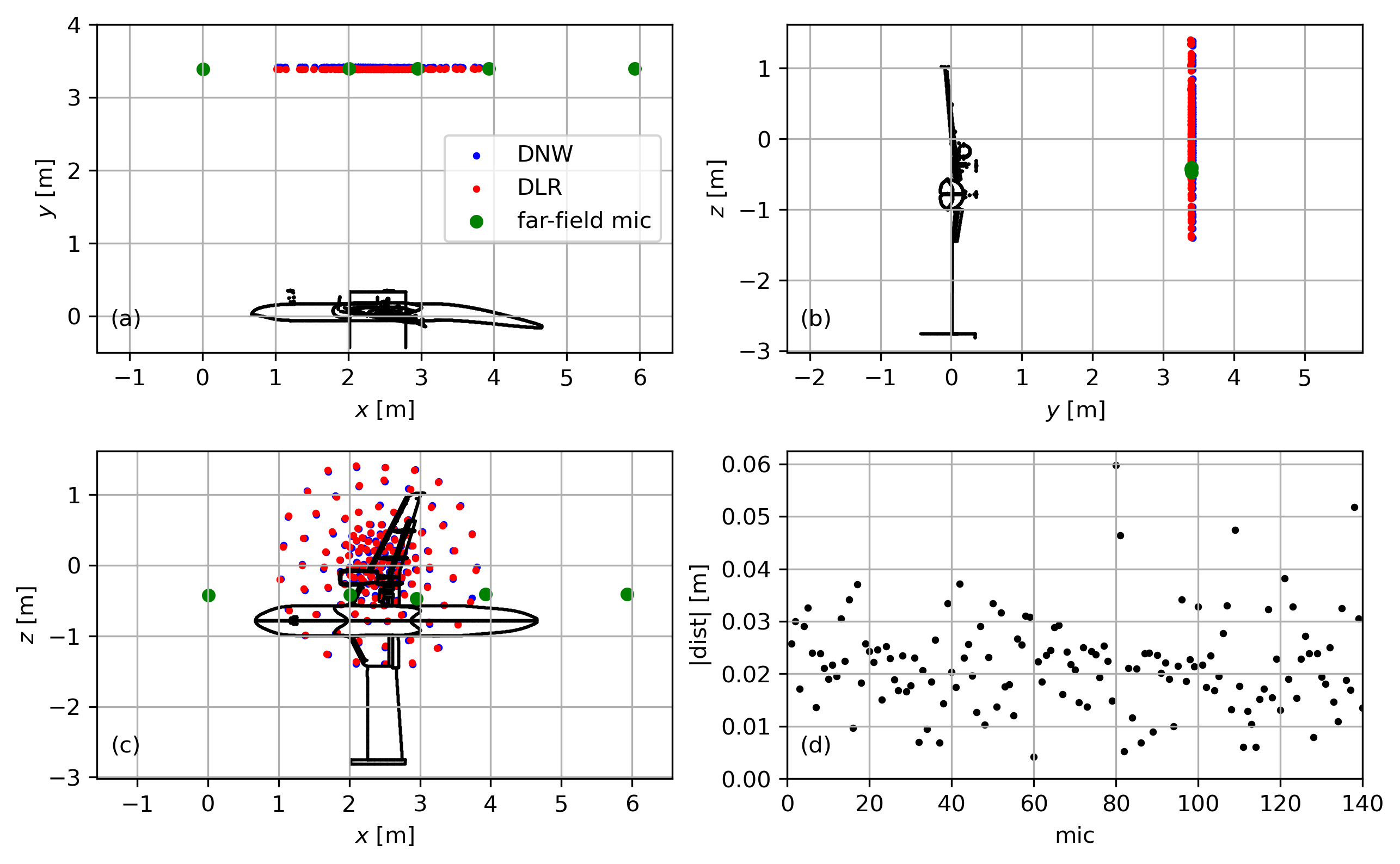}
\caption{$(a)$, $(b)$, and $(c)$ show projections of the experimental setup with the DNW array, and close MEMS positions from the DLR array. Additionally, the positions of far-field microphones are shown. $(d)$ shows the distance between the DNW and MEMS array positions.}
\label{fig:array_geometry}
\end{figure}

To compare the quality of single sensor signals, DNW and DLR microphones are compared to five far-field microphones, displayed in Figure~\ref{fig:array_geometry} (green). For each of the far-field microphones, the closest sensor of both the DLR and DNW array is chosen (not shown in the figure). For the DNW array, no microphone in the vicinity of the first and last far-field microphone is available. 

\begin{figure}[ht!]
\centering
\includegraphics[width=1.0\textwidth]{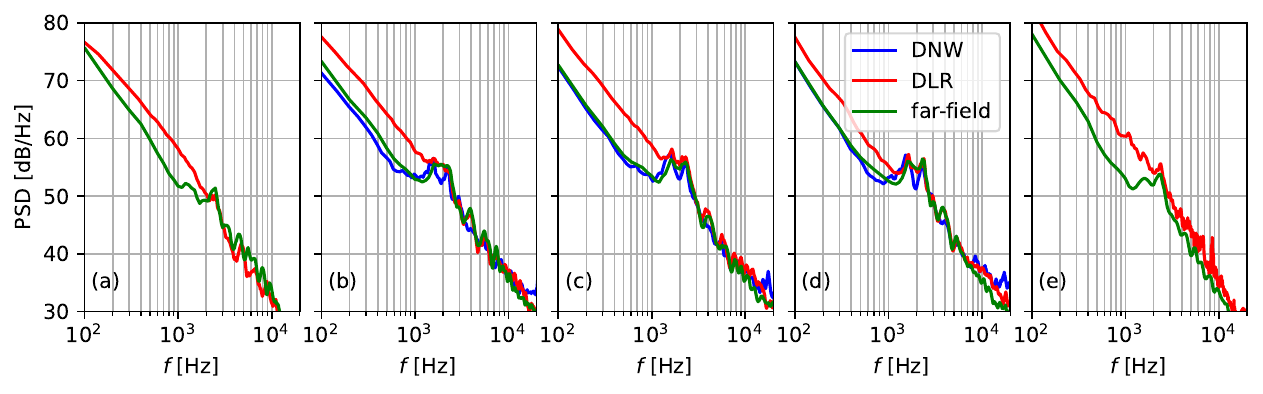}
\caption{$(a)$ to $(e)$ show the comparison of single sensor spectra for the array sensors closest to the corresponding far-field microphone in Figure~\ref{fig:array_geometry}. For $(a)$ and $(e)$ no DNW microphone is close enough for a comparison.}
\label{fig:spectra_farfield_compare}
\end{figure}

Figure~\ref{fig:spectra_farfield_compare} shows the corresponding spectra for these five positions. There is an excellent agreement between the spectra from $f\ge\SI{1500}{\hertz}$. Below this frequency, the DLR microphones show levels elevated up to $\Delta \text{PSD}\le \SI{8}{\decibel}$. The reason for these elevated levels is the missing windshield for the MEMS, as opposed to the shielded far-field and DNW microphones, and the strong wind recirculation. There is some variation in the peak frequencies, due to small differences in the flow speed between both test runs.

\begin{figure}[ht!]
\centering
\includegraphics[width=1.0\textwidth]{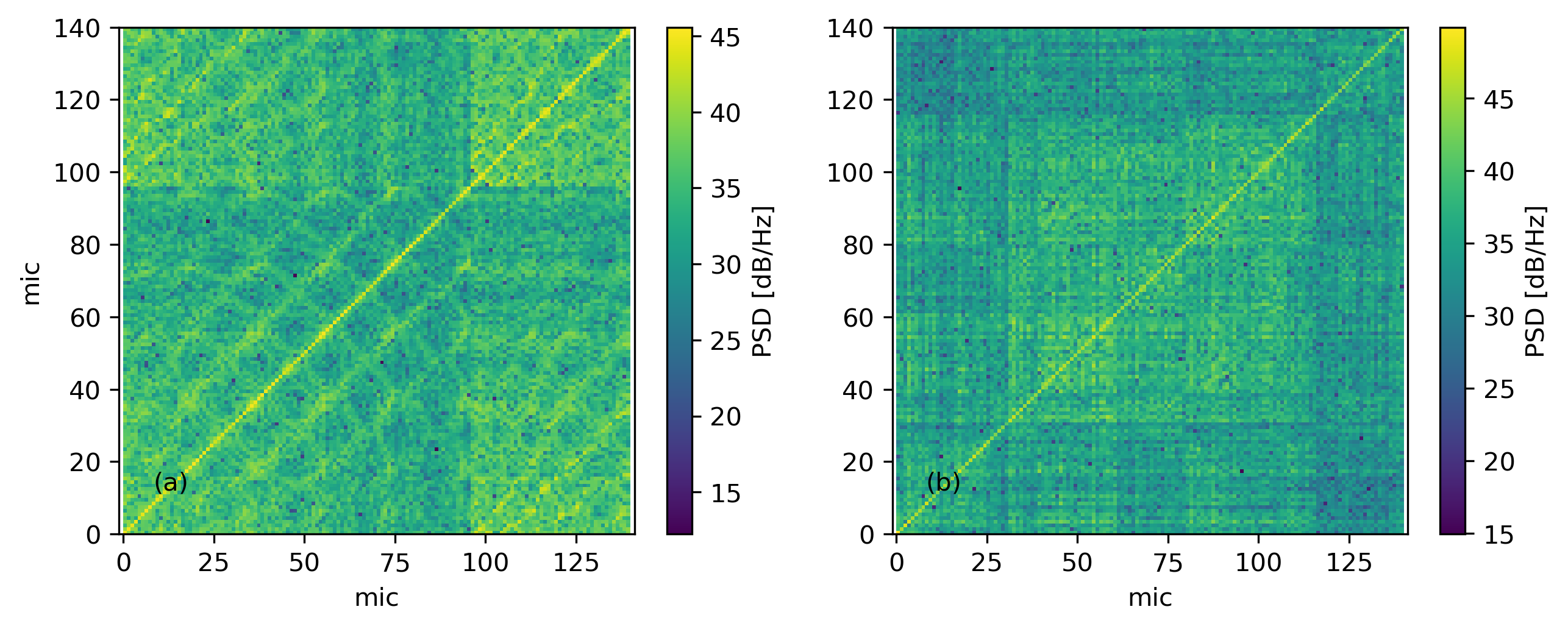}
\caption{Comparison of the $(a)$ DNW and $(b)$ DLR CSMs at $f=\SI{4}{\kilo\hertz}$.}
\label{fig:CSM_vgl_4kHz}
\end{figure}

To compare the performance of both arrays, their aperture, the number of sensors, and the sensor spacing must be similar. Since the DLR array features a large number of sensors, a DNW-like sub-array of 140 microphones was obtained from the DLR sensors. Figure~\ref{fig:array_geometry} shows the DNW array geometry and the corresponding DNW-like DLR geometry. Figure~\ref{fig:array_geometry} $(d)$ shows the distance for each sensor between both arrays, which is around $\SI{0.02}{\metre}$. 

\begin{figure}[ht!]
\centering
\includegraphics[width=1.0\textwidth]{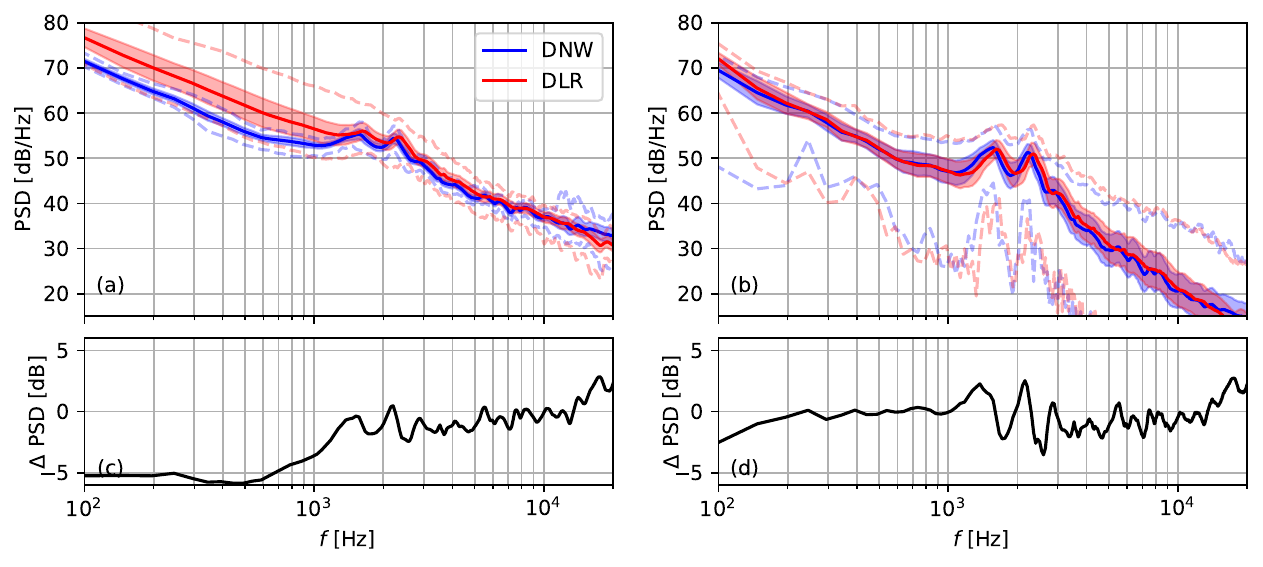}
\caption{$(a)$ shows the CSM auto-spectra averaged (full line), $1\sigma$ standard deviation (shaded area), and minimum and maximum (dashed lines) for the DNW and DLR CSMs. $(b)$ shows the corresponding cross-spectra. $(c)$, and $(d)$ show the difference, i.e. $\Delta \text{PSD}=\text{PSD}_\text{DNW}-\text{PSD}_\text{DLR}$.}
\label{fig:CSM_spectra_vgl}
\end{figure}

Figure~\ref{fig:CSM_spectra_vgl} shows the resulting CSM for the exemplary frequency $f=\SI{4}{\kilo\hertz}$. For the DNW array, a clear pattern based on the different Viper-acquisition modules is visible (each module has 8 channels). Further, there is a diagonal pattern visible of unknown origin.

\begin{figure}[ht!]
\centering
\includegraphics[width=1.0\textwidth]{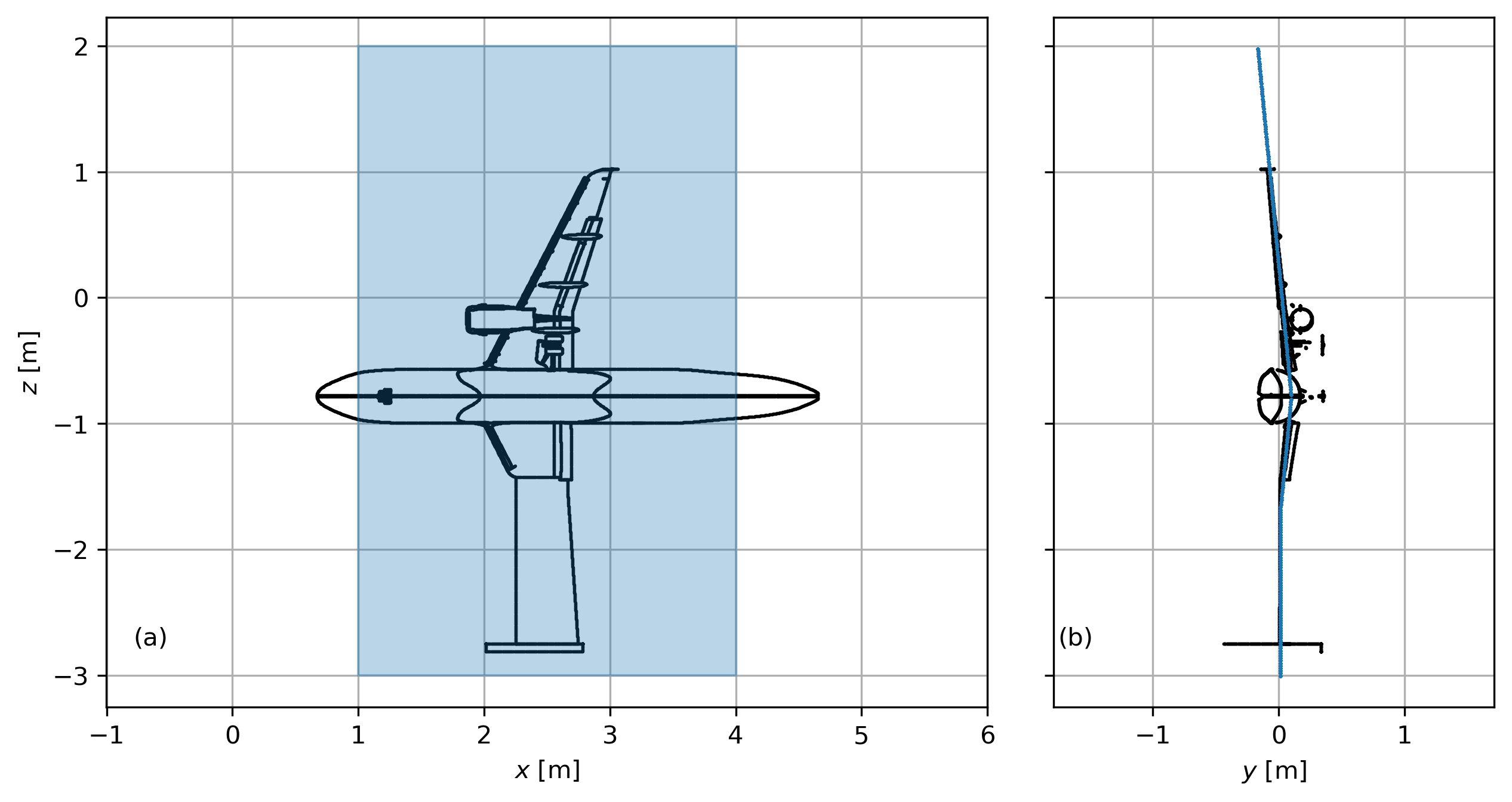}
\caption{2D focus grid for beamforming with $\SI{1}{\metre}\le x \le \SI{4}{\metre}, y\approx\SI{0}{\metre}, \SI{-3}{\metre}\le z \le \SI{2}{\metre}, \Delta x = \Delta z = \SI{0.02}{\metre}$ for a total of 37500 focus points. $(a)$ shows the ($x,z$)-projection, and $(b)$ shows the ($y,z$)-projection.}
\label{fig:bfm_grid}
\end{figure}

Based on all sensors of these comparable arrays, the signals are evaluated statistically. Figure~\ref{fig:CSM_spectra_vgl} shows the mean, standard deviation, minimum, and maximum of the $(a)$ auto-spectra (i.e. the CSM diagonal), and the $(b)$ cross-spectra (upper triangular CSM). While all auto-spectra statistically show the elevated levels below $f\le\SI{1500}{\hertz}$ in Figure~\ref{fig:CSM_spectra_vgl} $(c)$, the cross-spectra do not show this phenomenon in Figure~\ref{fig:CSM_spectra_vgl} $(d)$, since the recirculation wind-induced low-frequency noise is incoherent. Thus, it is not important for beamforming, as long as it does not exceed the cross-spectra's SNR (which depends on the number of block averages and thus, the measurement time). 

\begin{figure}[ht!]
\centering
\includegraphics[width=1.0\textwidth]{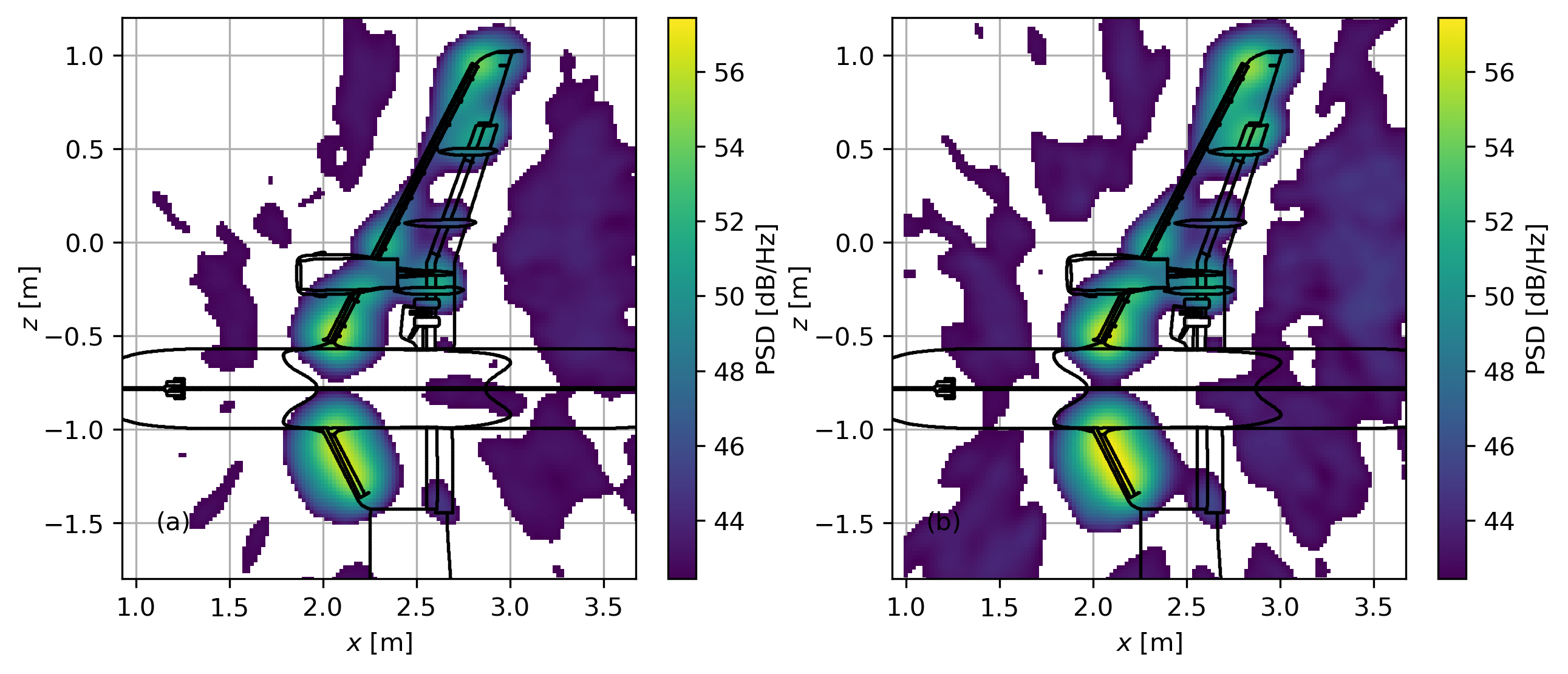}
\caption{Comparison of conventional beamforming maps with the $(a)$ DNW (140 microphones) and $(b)$ DLR array (140 MEMS microphones) at $f=\SI{4}{\kilo\hertz}$.}
\label{fig:conv_2D_vgl_f3rd_04000Hz}
\end{figure}

To compare the performance of both arrays beamforming is performed on an equidistant 2D focus grid, that follows the wing's delta angle, and the AoA. Figure~\ref{fig:bfm_grid} shows the focus grid, which covers the full wing. Conventional beamforming and CLEAN-SC is performed with steering-vector III~\cite{Sarradj2012}.

\begin{figure}[ht!]
\centering
\includegraphics[width=1.0\textwidth]{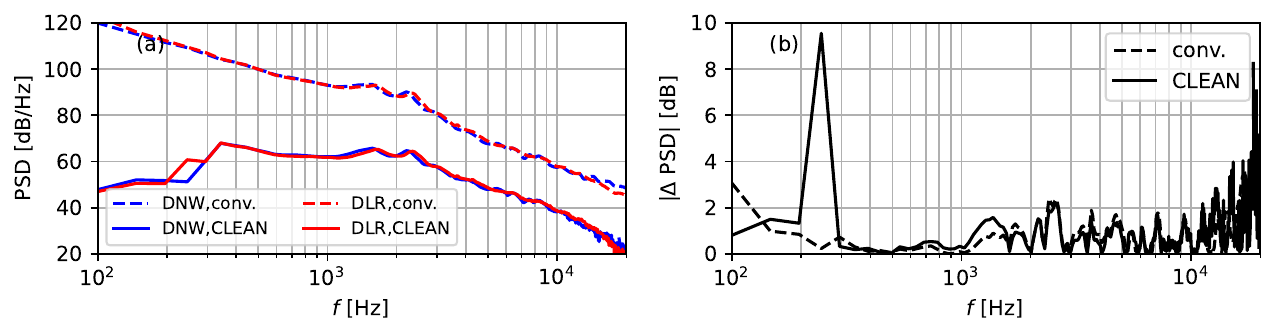}
\caption{$(a)$ shows the spectra resulting from spatially integrating the conventional (dashed line) and CLEAN-SC (full line) beamforming map. $(b)$ shows the absolute difference between the DNW and DLR array results.}
\label{fig:ROI_vgl_spectrum}
\end{figure}

Figure~\ref{fig:conv_2D_vgl_f3rd_04000Hz} shows an exemplary conventional beamforming map for $f_{3\text{rd}}=\SI{4}{\kilo\hertz}$, $(a)$ with the DNW array, and $(b)$ with the DLR array. Both maps are nearly identical, and all sources are equally well resolved. The DLR map shows slightly elevated levels at the bottom slat.

To compare the performance of the arrays for all frequencies the beamforming maps are spatially integrated around the upper wing, which results in spectra. Figure~\ref{fig:ROI_vgl_spectrum} $(a)$ shows the spectra for both the integrated conventional and CLEAN-SC beamforming maps. They are nearly identical, highlighted by their difference in Figure~\ref{fig:ROI_vgl_spectrum} $(b)$, which only shows a periodical error, based on the shifted peak frequencies due to the slightly different flow speeds.

These results highlight that the beamforming performance of both arrays is identical, given the similar geometry. This is true, even though the MEMS show elevated levels at low frequencies due to wind noise. Since the wind noise is uncorrelated, it does not influence the beamforming results.

\subsection{Directivity}\label{sec:results:directivity}

\begin{figure}[ht!]
\centering
\includegraphics[width=1.0\textwidth]{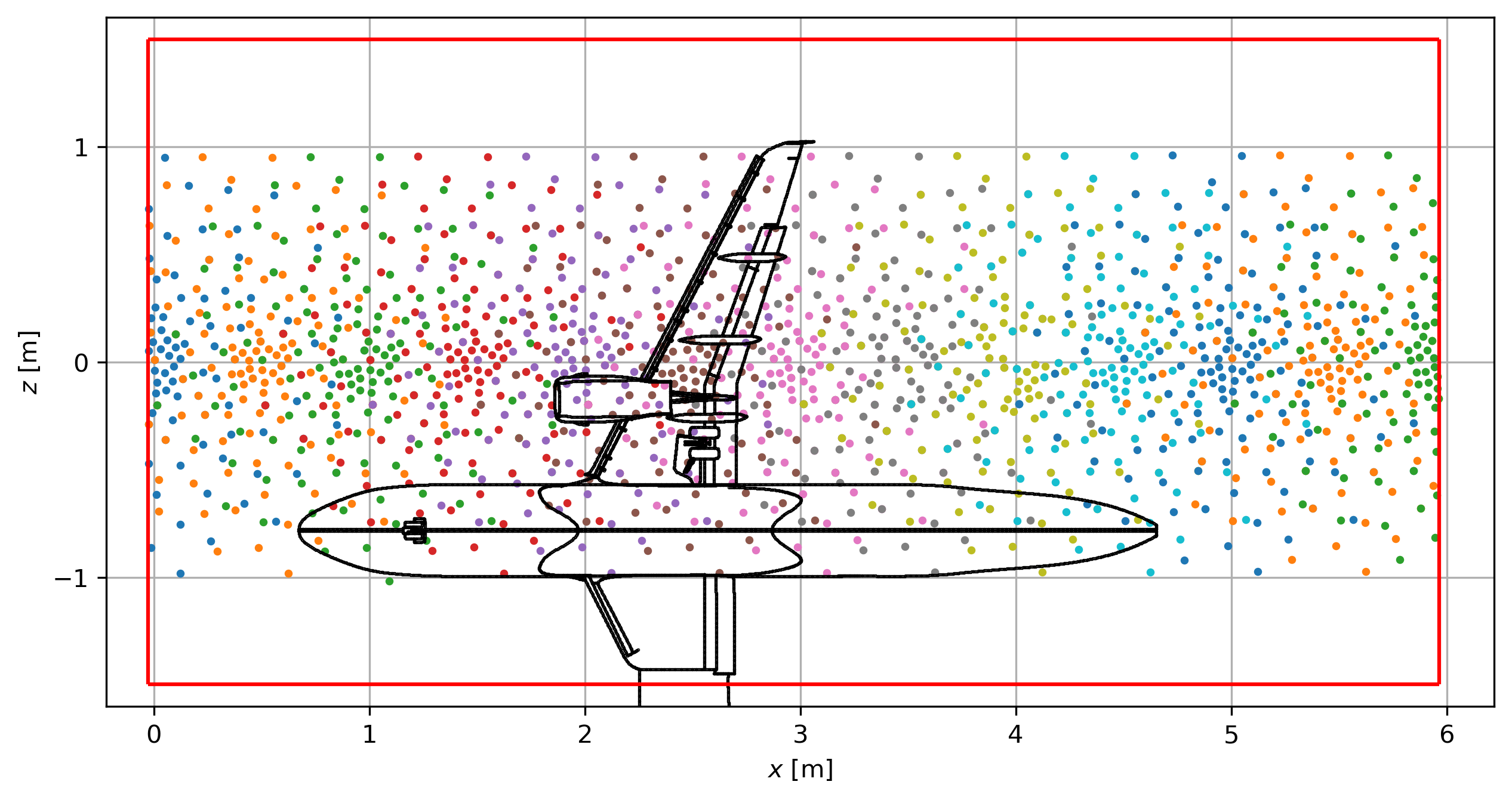}
\caption{$D=\SI{2}{\metre}$ aperture sub-arrays with up to $M=150$ microphones at various pitch angles $\theta$. The total array area is shown as a red box, and the various sub-array microphone positions with different colors.}
\label{fig:DLR_array_varib_x}
\end{figure}

This subsection presents a first assessment of pitch angle source directivity. Previously, the array had to be relocated to different positions, to assess sources from different angles~\cite{Ahlefeldt2018}. Due to the large size of the MEMS array, multiple sub-arrays can be simply sampled from the sensors to achieve the same result. The sampling of sensors is performed iteratively, based on an optimal array design (i.e., a Fermat spiral in this case). The optimal positions are calculated, and then for each optimal position, a sensor position is identified. If the sensor position is further away than a specified maximum distance $\varepsilon$, the optimal position is discarded. If a sensor position is assigned to an optimal position, it is removed from the pool of remaining sensor positions. This way each sensor position is only used once, and it is ensured that the array geometry is as close to the optimal geometry as possible, given the manually determined maximum distance $\varepsilon$, for which we use $\varepsilon=\SI{0.1}{\metre}$ for this application. 

\begin{figure}[ht!]
\centering
\includegraphics[width=1.0\textwidth]{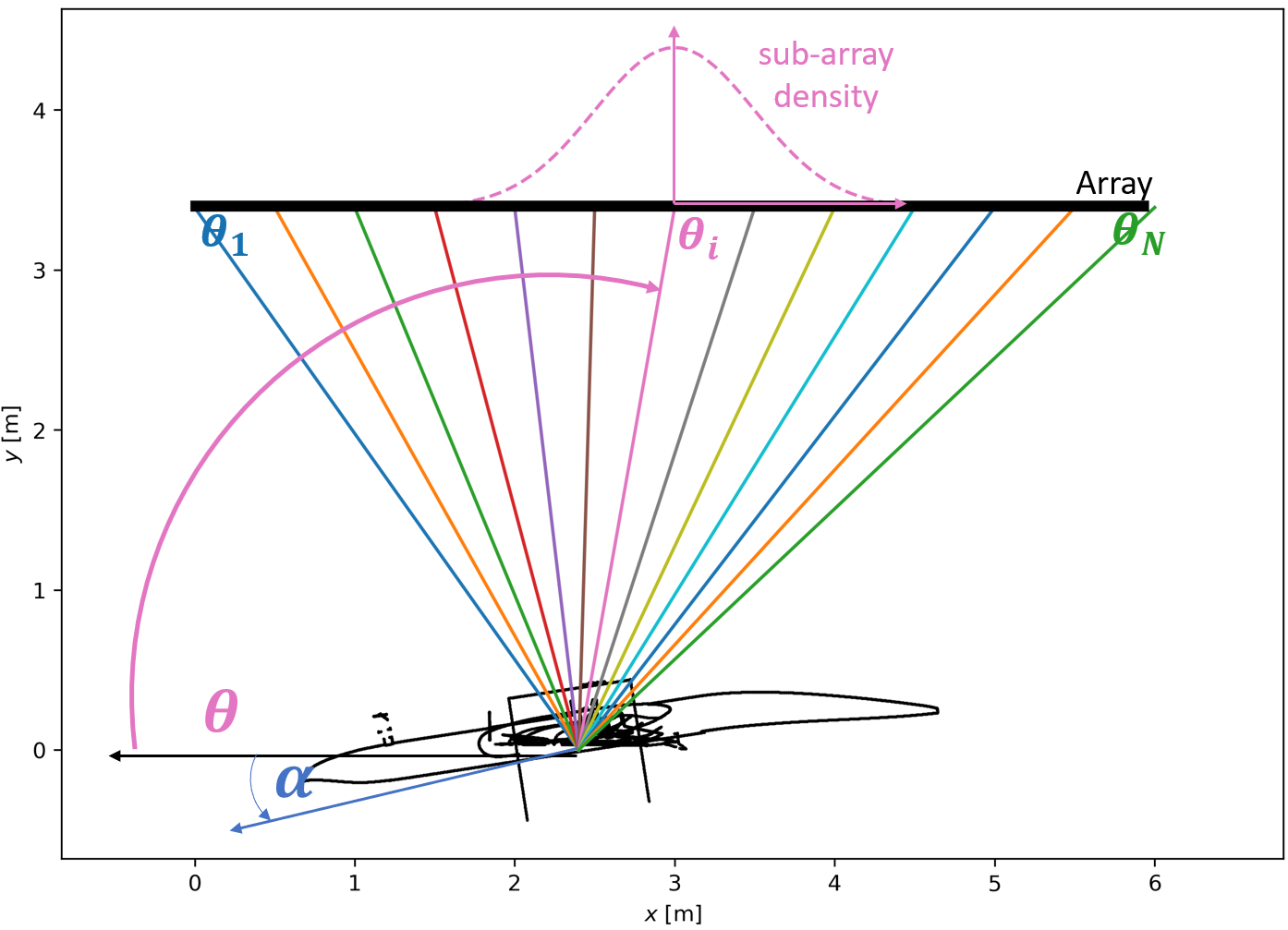}
\caption{Different observation angles $\theta$ for the sub-arrays. The observation angle is in reference to the model's rotation point and ignores its angle of attack. In reality, the observation angle is an area-averaged observation, based on the sub-array's spatial distribution and density.}
\label{fig:DLR_array_varib_y_explenation}
\end{figure}

Figure~\ref{fig:DLR_array_varib_x} shows the resulting sub-arrays for thirteen increasing pitch angles, based on a $M=150$ microphone Fermat spiral. Note, that the actual number of microphones is less for the first and last sub-arrays, due to the array boundary. Further, the array acts as a spatial filter, thus the actual directivity is the integration of a continuous directivity function, multiplied by the sub-arrays spatial filter. The average directivity angle can be assessed based on the geometric mean of the sensor positions of the sub-array. Additionally, the standard deviations give an estimate, of how much the observed directivity is smeared out due to the spatial filter of the array. Thus, the actual directivity (and PSF) will be distorted at the borders of the total array.

For example, the fifth sub-array has a nominal position of $x=\SI{2.500}{\metre}$, but the geometric mean is $\tilde{x}=\SI{2.598}{\metre}\pm \SI{0.362}{\metre}$. Thus, the nominal observation angle for a source at $x=\SI{2.400}{\metre}$ is $\theta = \SI{91.689}{\degree}$, and the true geometric observation angle is $\tilde{\theta} = \SI{93.338}{\degree}\pm\SI{6.082}{\degree}$.

\begin{figure}[ht!]
\centering
\includegraphics[width=1.0\textwidth]{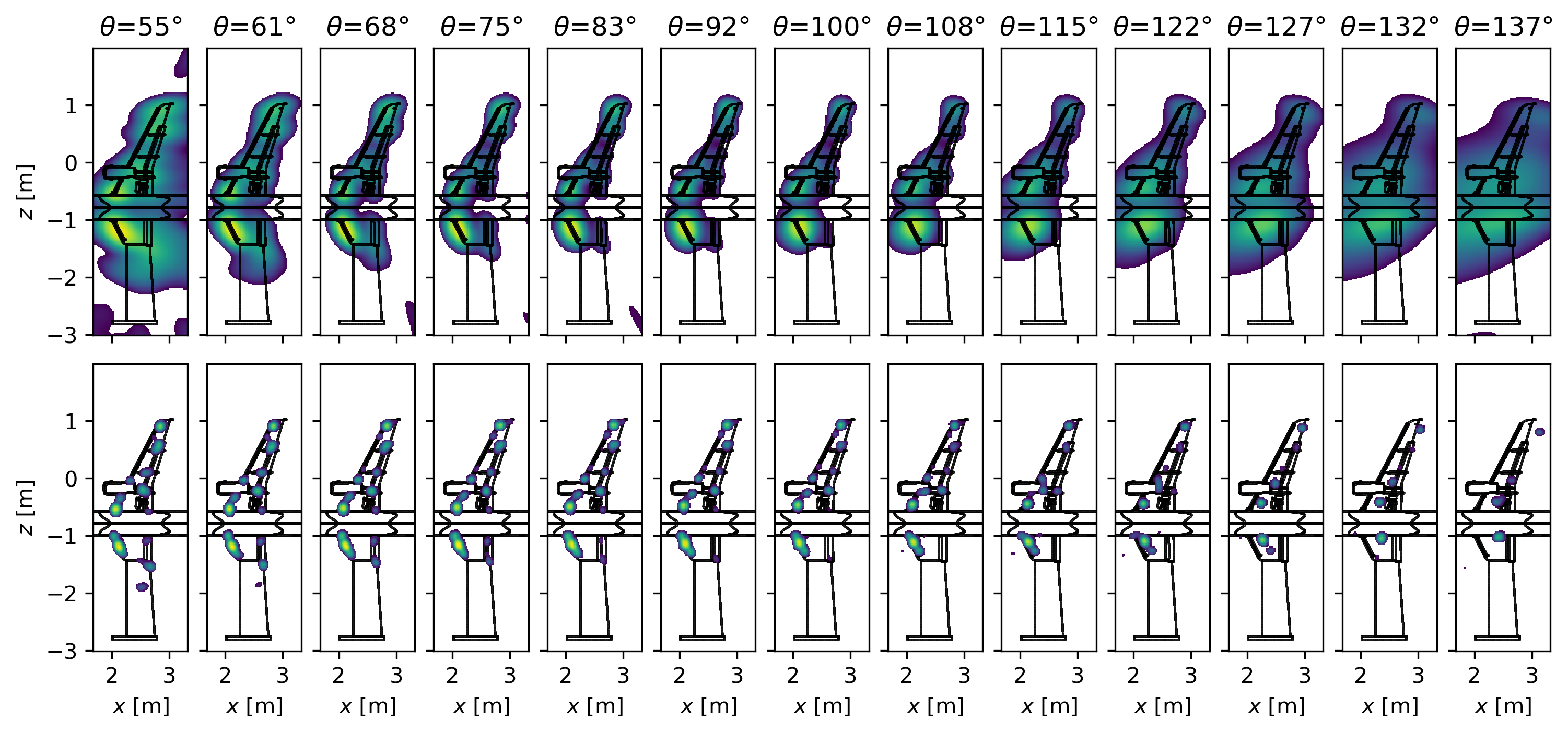}
\caption{Top: Conventional beamforming result; bottom CLEAN-SC result; both for $f=\SI{4}{\kilo\hertz}$. The dynamic range $\text{SNR}=\SI{15}{\decibel}$ for both rows is shared across all observation angles.}
\label{fig:direc_2D_vgl_f3rd_DLROPT01_04000Hz}
\end{figure}

Figure~\ref{fig:direc_2D_vgl_f3rd_DLROPT01_04000Hz} shows beamforming results at $f_{3\text{rd}}=\SI{4}{\kilo\hertz}$ for the sub-arrays with nominal observation angles $\SI{55}{\degree}\le\theta\le\SI{137}{\degree}$, the top row shows conventional beamforming results, the bottom row shows CLEAN-SC results (convoluted with a Gaussian kernel for better visibility). The Figure shows that at increasing downstream angles the resolution and level decrease. Further, the sources appear to move downstream with increasing angles in the CLEAN-SC maps. This is not due to directivity, but due to the shear layer refraction and scattering~\cite{ErnstSpehrBerkefeld2015}\cite{Ernst2020}. The Amiet open wind tunnel correction only assumes refraction along a plane, which is violated by the increasing shear layer thickness at downstream angles. The scattering results in a signal decorrelation, that depends on the thickness of the shear layer, the angle of incident, the frequency, and the distance between the sensor pairs. The decorrelation-induced lower coherence levels then result in an underestimation of the source levels with beamforming, especially at high frequencies. 

\begin{figure}[ht!]
\centering
\includegraphics[width=1.0\textwidth]{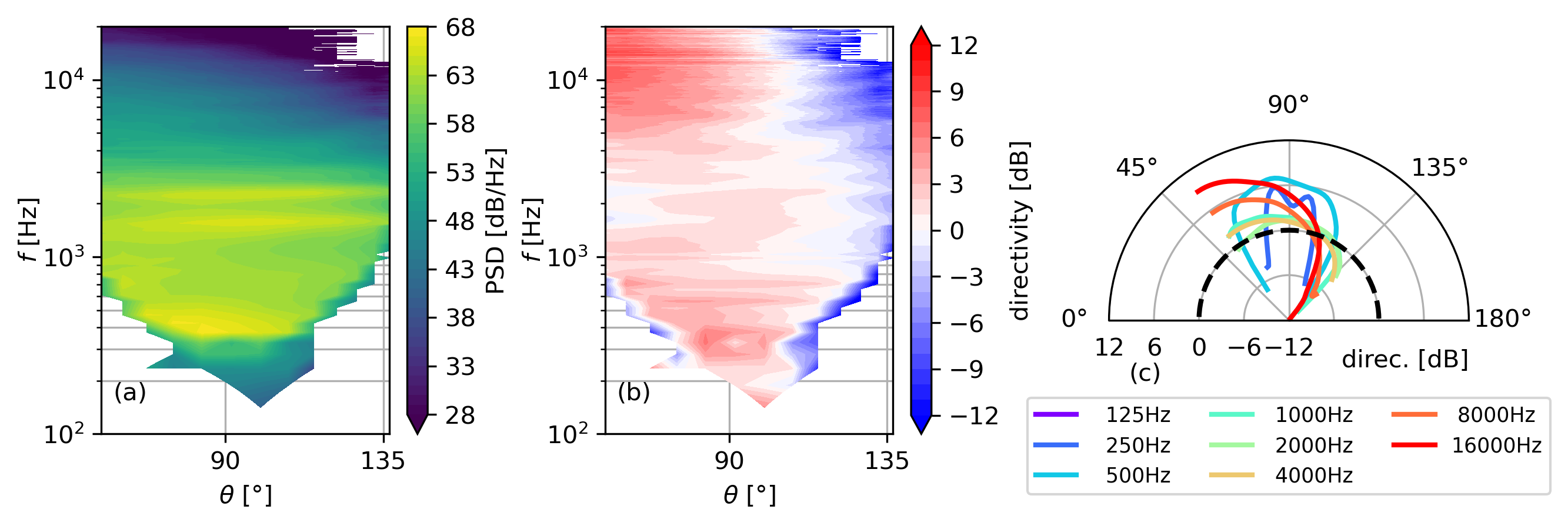}
\caption{$(a)$ shows the $\text{PSD}(\theta,f)$ based on the spatially integrated CLEAN-SC maps for the various sub-arrays from Figure~\ref{fig:DLR_array_varib_y_explenation}. $(b)$ shows the resulting directivity $\Gamma=\text{PSD}(\theta,f)-\langle\text{PSD}(\theta,f)\rangle_\theta$. $(c)$ shows the directivity in a polar plot for integrated octaves.}
\label{fig:DLR_varib_x_direc_02}
\end{figure}

Figure~\ref{fig:DLR_varib_x_direc_02} $(a)$ shows the spectra from the spatially integrated CLEAN-SC maps for all observation angles $\text{PSD}(\theta,f)$. The PSD shows the two prominent slat cove tones~\cite{Goudarzi2022} at $f=\SI{1500}{\hertz}$ and $f=\SI{2500}{\hertz}$, dominant at $\theta\approx\SI{90}{\degree}$. Above these frequencies, the level decreases with increasing frequency, and increasing observation angles. To obtain the source directivity at each frequency, the angle-average is subtracted from the PSD for each frequency, i.e. $\Gamma=\text{PSD}(\theta,f)-\langle\text{PSD}(\theta,f)\rangle_\theta$. Thus, the directivity $\Gamma=\SI{0}{\decibel}$ indicates, that at the given angle the average source power is observed. $\Gamma\ge\SI{0}{\decibel}$ indicates that the source radiates stronger towards the angle, and $\Gamma\le\SI{0}{\decibel}$ indicates that the source radiates less energy into the given direction. Figure~\ref{fig:DLR_varib_x_direc_02} $(b)$ shows this directivity plot, and $(c)$ shows the directivity for integrated octaves in a polar plot. Up to $f\le\SI{3}{\kilo\hertz}$ most of the energy is radiated towards $\theta\approx\SI{90}{\degree}$, with increasing frequency the energy is radiated upstream. Note again, that this effect is mostly driven by the shear layer induced decorrelation.

\subsection{Beamforming to far-field projection}\label{sec:results:farfield}
\begin{figure}[ht!]
\centering
\includegraphics[width=1.0\textwidth]{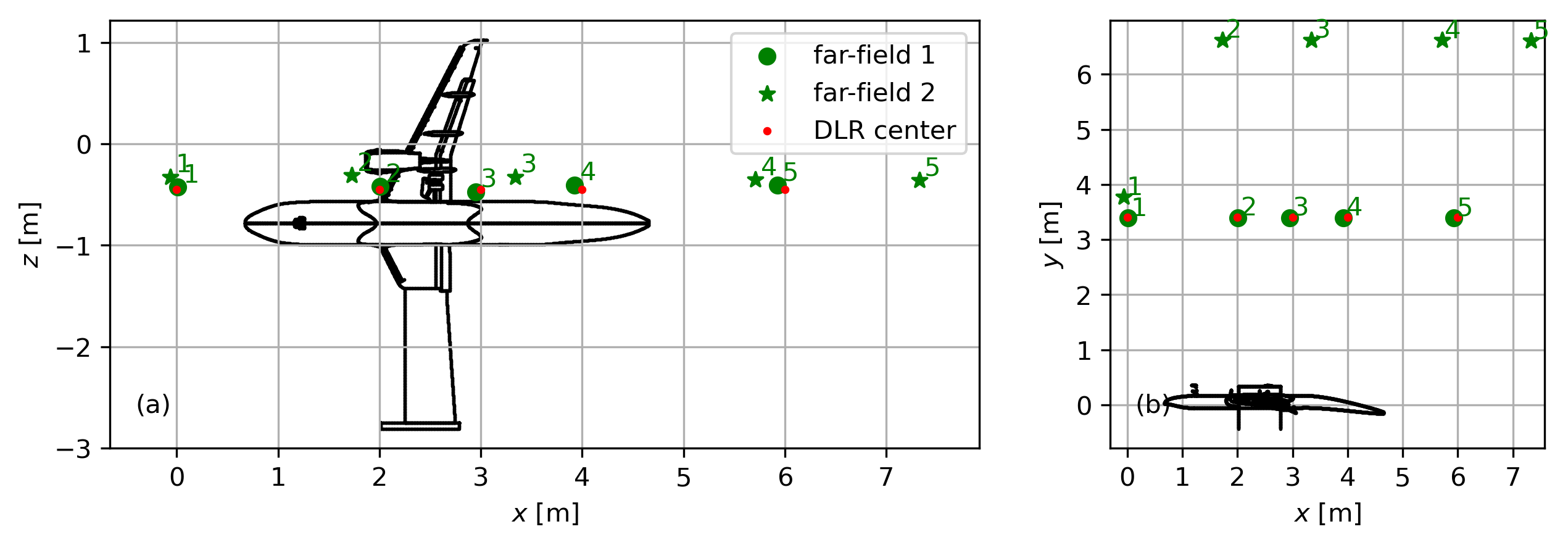}
\caption{$(a)$ and $(b)$ show projections of two far-field microphones (dots, and stars), that lie along the same observation angle. The corresponding DLR sub-array centers are marked.}
\label{fig:farfield_VS_direc_geom}
\end{figure}

This subsection assesses the ability of beamforming to predict far-field levels in the open test section. For this, two rows of far-field microphones are compared to beamforming results at the given observation angles. Figure~\ref{fig:farfield_VS_direc_geom} shows the locations of the microphones and the centers of the DLR sub-arrays. The two rows of far-field microphones are spaced at two different distances from the model, but at the same angle of incident (except microphone 1, row 2, due to spatial limitations). To compare the results of beamforming, which estimates the source power with a monopole assumption, and the far-field microphone, which measures the sound immission at a given distance, both predictions must be converted to the same unit. To do so, the beamforming result is calculated as the sound pressure that would be observed at $d_0=\SI{1}{\metre}$ from the source. The far-field microphones are then corrected in their level, based on their distance to the source with $\Delta \text{PSD}=20\log_{10}(d/d_0)$, where $d_0=\SI{1}{\metre}$ is the reference distance of the beamforming result. 

\begin{figure}[ht!]
\centering
\includegraphics[width=1.0\textwidth]{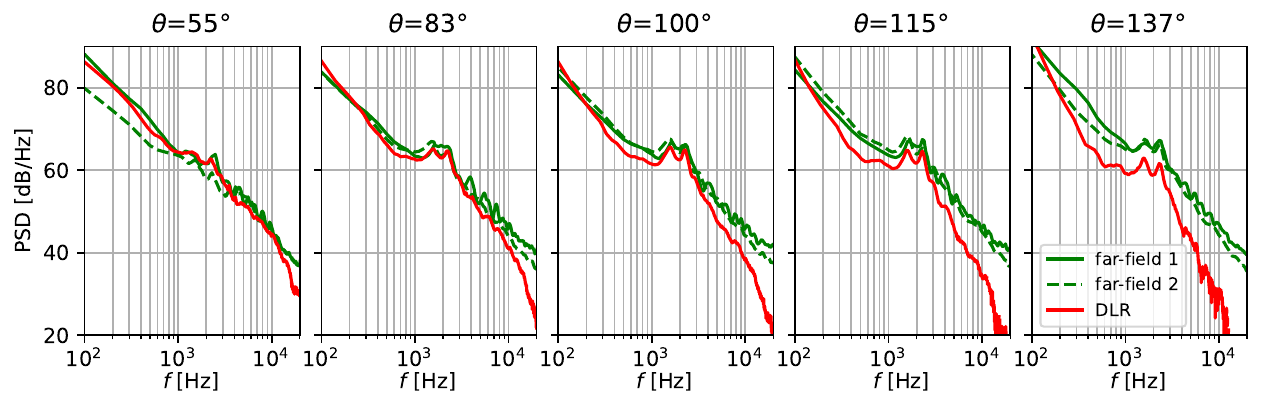}
\caption{Comparison of the five far-field microphone spectra at the observation angles in Figure~\ref{fig:farfield_VS_direc_geom} and the spatially integrated CLEAN-SC maps. The far-field spectra are distance normalized with $\Delta \text{PSD}=20\log_{10}(d)$, where $d$ is their distance to the model's rotation point.}
\label{fig:farfield_VS_direc_spectra}
\end{figure}

Figure~\ref{fig:farfield_VS_direc_spectra} shows the resulting spectra of the spatially integrated CLEAN-SC results and the distance-corrected far-field microphones. The spectra show a very good agreement at low frequencies and small observation angles. With increasing angles, beamforming underestimates the source power, and with increasing frequencies, this underestimation is amplified, due to the aforementioned shear layer decorrelation.

A common approach to counter this effect at high frequencies is to use smaller sub-arrays at increasing frequencies. Due to the fixed array geometry, this has been performed by applying a window function on the CSM. This window function weights CSM entries less (or not at all) with an increasing distance between sensor pairs, to counter the aforementioned coherence loss. However, this results in less available CSM entries and affects the dynamic range and resolution negatively. We adapt this approach by sampling a frequency-dependent sub-array from all available sensor positions. This frequency-dependent array uses a Fermat spiral with an aperture $D=\SI{5.5}{\metre}$ at $f=\SI{1}{\kilo\hertz}$ and up to $M=200$ microphones. Then, with increasing frequencies, the aperture is scaled accordingly with $D/f$ up to $f=\SI{16}{\kilo\hertz}$. For the resulting optimal Fermat arrays, the actual microphone positions are then iteratively estimated as described above.

\begin{figure}[ht!]
\centering
\includegraphics[width=1.0\textwidth]{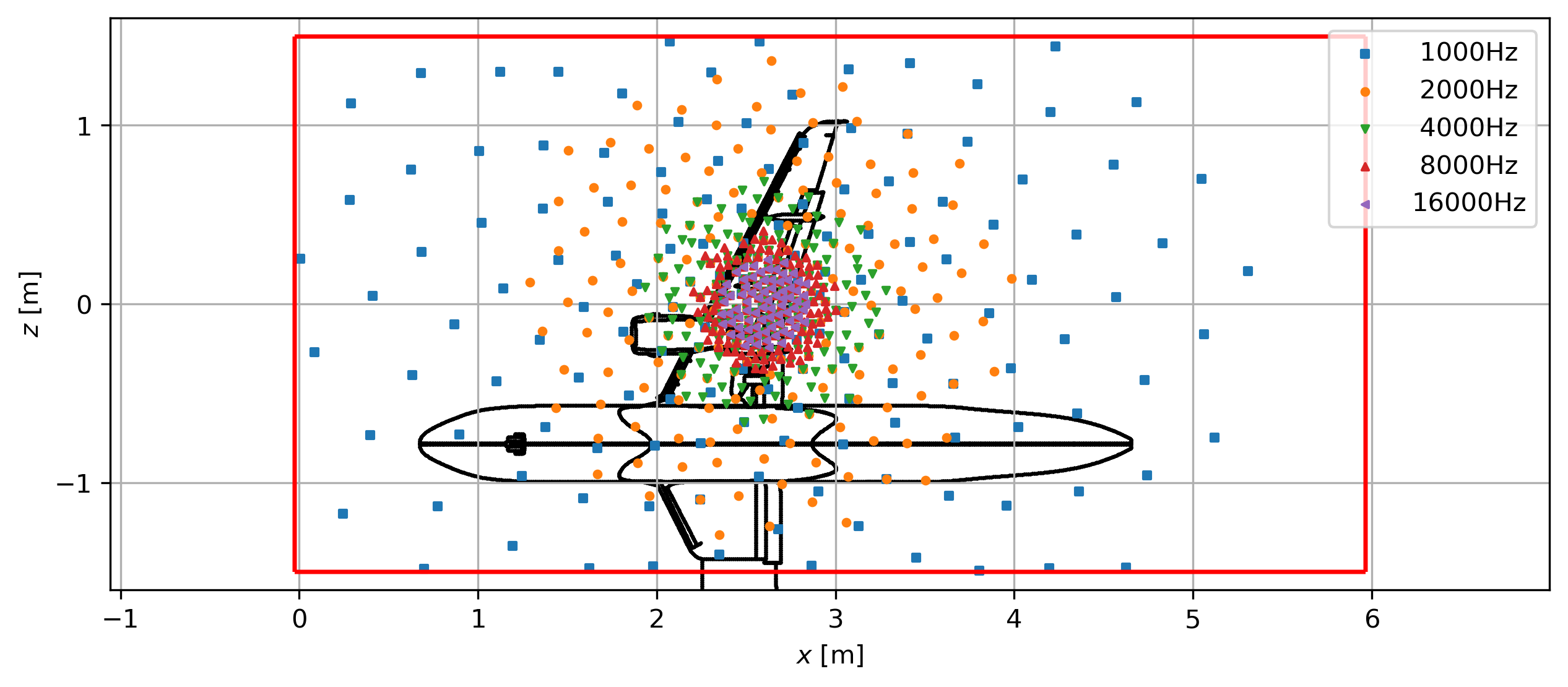}
\caption{Frequency-dependent sub-array geometry at $\theta\approx\SI{100}{\degree}$, where the aperture shrinks linearly with increasing frequency, shown for exemplary frequencies. }
\label{fig:DLR_array_varib}
\end{figure}

Figure~\ref{fig:DLR_array_varib} shows sub-array geometry at $\theta=\SI{100}{\degree}$ for five different frequencies. Note, that from $f\ge\SI{4}{\kilo\hertz}$ the resulting sub-array starts to reassemble a circular form instead of the desired Fermat spiral, based on the available microphones within the given vicinity. In theory, additional shading can be applied to the resulting sub-arrays, but for this study, the aperture is only varied using different physical sensors. Note, that this frequency-dependent aperture is optimal for beamforming: At low frequencies sources typically do not have strong directivities, so a large aperture is not problematic for their assessment. Further, decorrelation effects are weak, so that beamforming profits from the large aperture because of the increased spatial resolution. At increasing frequencies, the resolution is kept constant with a shrinking aperture. At the same time, the decorrelation effect countered to some degree and the smaller aperture allows for a more detailed assessment of the source directivity. 

\begin{figure}[ht!]
\centering
\includegraphics[width=1.0\textwidth]{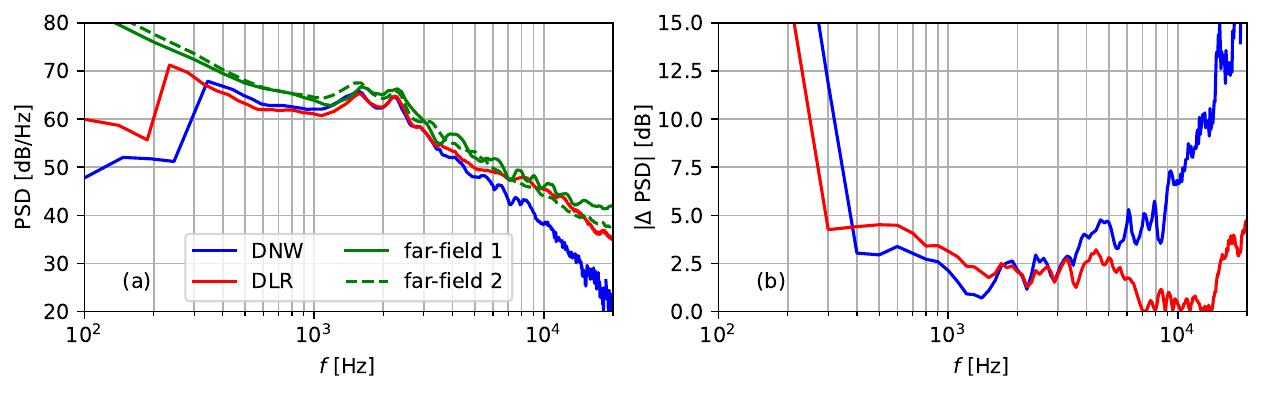}
\caption{$(a)$ shows the spatially integrated CLEAN-SC maps from the standard DNW array and the frequency-dependent sub-arrays, see Figure~\ref{fig:DLR_array_varib}, and the corresponding distance normalized far-field spectra, see Figure~\ref{fig:farfield_VS_direc_geom}. $(b)$ shows the absolute difference between the CLEAN-SC spectra and the average far-field spectra for the DNW and DLR array.}
\label{fig:ROI_vgl_VARARRAY}
\end{figure}

Figure~\ref{fig:ROI_vgl_VARARRAY} $(a)$ shows the resulting spatially integrated CLEAN-SC spectra for both the standard DNW array (without shading), and the frequency-dependent aperture DLR array at $\theta=\SI{100}{\degree}$. All differences between both arrays are based on the changing aperture, since for the same array geometry the results were identical, see Figure~\ref{fig:ROI_vgl_spectrum}. Due to the increased aperture at low frequencies, the main lobe width of the DLR array is decreased by a factor of two compared to the DNW Array. At medium frequencies $\SI{400}{\hertz}\le f\le\SI{5}{\kilo\hertz}$ the results remain nearly identical. At increasing frequencies, the decreasing aperture of the DLR array completely counters the shear layer decorrelation. Figure~\ref{fig:ROI_vgl_VARARRAY} $(b)$ shows this as the difference between the DLR and DNW array spectra and the averaged far-field spectra. The variable DLR array has low deviations $\Delta \text{PSD}\le\SI{5}{decibel}$ from the ground truth far-field spectra for $\SI{300}{\hertz}\le f\le\SI{20}{\kilo\hertz}$. Thus, beamforming in the near-field with a monopole assumption and calculating the resulting far-field sound immission levels based on a monopole assumption is valid as the assumptions (though they are typically wrong for airframe noise) cancel out in this process. Further, beamforming can predict absolute levels in open wind tunnels and not only relative levels. 

\section{Outlook}\label{sec:outlook}

This paper presented the first use of a massively upscaled array and showed, that this has several advantages. In this paper, we showed that the large array can be used to derive sub-arrays, to improve the spatial localization of sources, or to derive source directivities. The array was tested in the open test section with success and will be used in productive tests, with further upscaling.

The array showed great improvements in the setup and data recording, based on the few cables and data acquisition systems necessary. Further improvements of the array should include a variety of sensors and actuators that go beyond acoustic data acquisition, such as temperature sensors for the estimation of the speed of sound, acoustic actuators for self-calibration, automatic detection of defect sensors, or advanced active source detection methods.

This paper presented a first analysis of the directivity of airframe noise using multiple sub-arrays. It was shown, that the source power prediction with beamforming is strongly affected by the loss of signal coherence due to refraction and scattering of the sound waves. While a frequency-dependant aperture array helped to counter this effect partially, a model to correct the loss of coherence based on the shear layer exists~\cite{Ernst2020} and has to be integrated into the steering vector, to estimate the source levels correctly. Further research is necessary if an improved model of the Amiet open wind tunnel correction for the shear layer at large downstream angles can account for the drift of the observed sources. Further, the presented directivity analysis was performed for the overall sound emission of the model. A detailed analysis of individual sources is necessary since the individual sources have different distances from the far-field microphones. To do so, the analysis should be performed with a detailed, 3D focus grid~\cite{Ahlefeldt2023}.

This paper showed that the MEMS array in combination with CLEAN-SC can predict the resulting far-field sound field with great accuracy in an open wind tunnel above the frequency threshold, where the array's aperture can no longer resolve the sources. This experiment must be repeated in a closed wind tunnel.

\section{Conclusion}\label{sec:conclusion}
This paper presented the successful design of a modular $\SI{1}{\metre}\times\SI{2}{\metre}$ MEMS array panel, that can be stacked to achieve a large array area without gaps. The array, consisting of $3\times3$ panels with a total of 7200 sensors was used to test its capabilities against a standard 140-microphone spiral array and far-field microphones on a 1:9.5 airframe half model in an open wind tunnel section at DNW-NWB. 

Individual MEMS sensors showed elevated levels at low frequencies due to the lack of wind shielding and a strong recirculation at the array's position, and a very good agreement at medium and high frequencies with the DNW microphones and far-field microphones. These elevated levels are only observed for the auto spectra, but not for the cross spectra so that they are irrelevant for beamforming with diagonal removal. A statistical analysis of a MEMS sub-array, chosen so it matches the DNW array's geometry, showed an identical performance of both arrays.

Conventional beamforming and CLEAN-SC were performed on a 2D grid that follows the shape of the wing and the angle of attack. The results were identical for both arrays, showing that the MEMS array is suitable for reproducing the state-of-the-art DNW array's results.

Sub-arrays at different angles of incident along the pitch angle were chosen for beamforming to obtain an estimation of the directivity of the model's source emission. The results showed, that the sound power estimation at increasing downstream angles is mostly dominated by the loss of coherence through shear layer decorrelation and refraction.

The CLEAN-SC maps were spatially integrated to obtain spectra. The CLEAN-SC results and the far-field microphones' spectra were then normalized to a common distance to show of beamforming can correctly estimate the resulting far-field sound field. The results showed very good agreements at upstream angles and up to medium frequencies. At increasing frequencies and increasing observation angles, the shear layer decorrelation resulted in an underestimation of the resulting sound field.

This was corrected by introducing a sub-array with a frequency-dependent aperture, similar to shading, but by actually changing the sensor layout with increasing frequency. This approach showed two positive effects on the results: First, due to the improved resolution at low frequencies beamforming was able to estimate the spectrum down to lower frequencies. Second, the shrinking aperture at high frequencies resulted in smaller sensor pair distances, which reduced the shear layer-induced loss of coherence. This resulted in the correct estimation of the source emission in the high-frequency range.

\section*{Author contributions}
\textbf{Project}: [Conceptualization: Ernst, Spehr; Funding acquisition: Spehr; Administration: Ernst] \textbf{Array development}: [FPGA: Geisler; Array design: Ernst; Calibration: Ahlefeldt, Ernst; Software: Philipp, Ernst] \textbf{Experiment}: [Ernst] \textbf{Results}: [Goudarzi] \textbf{Paper}: [Goudarzi]

\section*{Acknowledgments}
The measurements were carried out within the DLR project SIAM “Low Noise Medium-Range Aircraft”. Many people from the DLR were involved in this comprehensive project. We would like to thank them all, but cannot mention all of their names here. In particular, we would like to thank Jakob Faust for a GPU implementation of the data conversion, Carsten Fuchs and Tobias Herrmann for mechanical support, Tobias Kleindienst and Kevin Kienass for providing countless prototypes, and Michael Pott-Pollenske responsible for the airframe model. We also would like to thank the DNW-NWB wind tunnel facility for the provided infrastructure and help in performing the measurements.

\bibliography{main.bib}

\end{document}